\documentclass[twocolumn,pra,twocolumn,superscriptaddress]{revtex4}
\usepackage[latin9]{inputenc}
\usepackage{color}
\usepackage{amsmath}
\usepackage{amssymb}
\usepackage{graphicx}
\usepackage{tikz}
\usepackage{mathrsfs} 
\usepackage{float}
\usepackage{bbm}
\usepackage{epstopdf}
\usepackage{epsfig}
\usepackage{verbatim}
\usepackage{array}
\usepackage{setspace}
\usepackage{times}

\usepackage[unicode=true,
bookmarks=true,bookmarksnumbered=false,bookmarksopen=false,
breaklinks=false,pdfborder={0 0 1},backref=false,colorlinks=true]
{hyperref}
\hypersetup{
	linkcolor=magenta, urlcolor=blue, citecolor=blue, pdfstartview={FitH}, hyperfootnotes=false, unicode=true}

\newcolumntype{C}[1]{>{\centering\arraybackslash$}p{#1}<{$}}

\begin{document}

\title{Implementation of geometric quantum gates on microwave-driven  semiconductor charge qubits}

\author{Chengxian Zhang}

\author{Tao Chen}
\affiliation{Guangdong Provincial Key Laboratory of Quantum Engineering and Quantum Materials, 
	and School of Physics\\ and Telecommunication Engineering, South China Normal University, Guangzhou, Guangdong 510006, China}

\author{Xin Wang}\email{x.wang@cityu.edu.hk}
\affiliation{Department of Physics, City University of Hong Kong, Tat Chee Avenue, Kowloon, Hong Kong SAR, China}
\affiliation{City University of Hong Kong Shenzhen Research Institute, Shenzhen, Guangdong 518057, China}
\author{Zheng-Yuan Xue}  \email{zyxue@scnu.edu.cn}
\affiliation{Guangdong Provincial Key Laboratory of Quantum Engineering and Quantum Materials, 
	and School of Physics\\ and Telecommunication Engineering, South China Normal University, Guangzhou, Guangdong 510006, China}
\affiliation{Guangdong-Hong Kong Joint Laboratory of Quantum Matter, and Frontier Research Institute for Physics, South China Normal University, Guangzhou, Guangdong 510006, China}

\date{\today}

\begin{abstract}
A semiconductor-based  charge qubit, confined in double quantum dots, can be a platform to implement quantum computing. However, it suffers severely from charge noises. Here, we provide a theoretical framework to implement universal geometric quantum gates in this system. We find that, while the detuning noise can be suppressed by operating near its corresponding sweet spot, the tunneling noise, on the other hand, is amplified and becomes the dominant source of error for single-qubit gates, a fact previously insufficiently appreciated. We demonstrate, through numerical simulation,  that the geometric gates outperform the dynamical gates across a wide range of tunneling noise levels, making them particularly suitable to be implemented in conjunction with microwave driving. To obtain a nontrivial two-qubit gate, we introduce a hybrid system with charge qubits coupled by a superconducting resonator. When each charge qubit is in resonance with the resonator, it is possible to construct an entangling geometric gate with fidelity higher than that of the dynamical gate 
for experimentally relevant noise levels. Therefore, our results suggest that geometric quantum gates are powerful tools to achieve high-fidelity manipulation for the charge qubit.

\end{abstract}

\maketitle

\section{Introduction}
Semiconductor-based quantum-dot qubits are promising candidates to realize universal quantum computation. Much progress has been made using the quantum-dot devices based on the spin and charge degrees of freedom, the former of which can form various types of spin qubits \cite{Loss.98,DiVincenzo.00,Petta.05,Laird.10,Medford.13,Huang.19b} while the latter can be used to encode the charge qubits \cite{Shinkai.09,Petersson.10,Li.15,Kim.15, Ward.16,Yang.19b,Pomorski1.19,Pomorski.20a,Pomorski.20b} by confining one electron in semiconductor quantum dot \cite{fujisawas.04,likharev.99}. Although there is a vast success of silicon-based spin qubits,
it is difficult to operate the oscillating (microwave) magnetic field and the gate operations there are slow \cite{Huang.19b},  
which shadows the prospect for the scalability. On the contrary, the charge qubit is fully controllable by gate voltages, making it easy to apply microwave pulses, and thus leading to fast quantum gates  \cite{Kim.15}. 	
However, high-fidelity manipulation of charge qubits remains challenging as they are severely affected by charge noises. Several theoretical approaches have been proposed to mitigate the noise effects and improve the gate fidelities, including, for example, pulse engineering \cite{Emerson.19} and dynamically corrected gates \cite{Wang.12,Kestner.13,Wang.14a,Wang.14b,Throckmorton.17}.  Recently, it has been demonstrated that strongly microwave-driven operations near the detuning sweet spots can effectively suppress the charge noises \cite{Kim.15,Wong.16,Nichol.17,Yang.17,Yang.19a,Yang.19b}. Despite these progresses, experimental gate fidelities are still below 90\% \cite{Kim.15} owing to the residual charge noise. Therefore, further improving the gate fidelities and enhancing their robustness against noises is key to the realization of quantum computation with semiconductor charge qubits.

Dynamically corrected gates and other related schemes focus on suppressing noises in the dynamical process, which is sensitive to local noise fluctuations. In 1984, Berry found that after the cyclic and adiabatic evolution, the quantum state can obtain an extra phase factor (Abelian geometric phase) having global property, i.e. it is determined only by the closed path of a cyclic evolution and is therefore robust to most local noises \cite{Berry.84}. Inspired by Berry's idea, quantum gates implemented with geometric phases \cite{Berry.84,Wilczek.84,Aharonov.87}, called geometric quantum gates (GQGs), are proposed to realize high-fidelity quantum computation \cite{Pachos.99,Zanardi.99,Duan.01}. Since then, much attention has been paid to GQGs, which can be realized based on either the Abelian geometric phase or the non-Abelian geometric phase. The GQGs based on the Abelian geometric phase has been realized in various system with two-level system \cite{Xiang.01,Li.02,Zhu.02,Zhu.03,Solinas.03a, Wang.16,Zhao.17,Tao.18,Xu.19,Zhang.20}. Different from the Abelian geometric phase, the non-Abelian geometric phase is related to the matrix-valued geometric phase (or quantum holonomy). The non-Abelian structure or the holonomy appears when more than one state is considered. The key for  the non-Abelian phase is that the subspace of the full state space needs to trace out a loop in the whole space \cite{Sjoqvist.15}. Since the non-Abelian GQGs are using the state space more than two-level structure, it is usually implemented in the three-level system \cite{Solinas.03b,Feng.13,AbdumalikovJr.13,Arroyo.14,Zu.14,Yale.16,Sekiguchi.17,Li.17, Xu.18,hong.18, Zhou.18, Shkolnikov.18, Shkolnikov.19,Egger.19,Yan.19, Zhu.19, ai.20}. On the other hand, both the Abelian and non-Abelian GQGs can be implemented by using the adiabatic \cite{Pachos.99,Zanardi.99,Falci.00,Wu.13,Toyoda.13,Huang.19geo,Frees.19} or non-adiabatic \cite{Erik.12,Xu.12} evolution of quantum states. The non-adiabatic GQGs have attracted much more interest compared to the adiabatic approach, as the latter requires an overly long gating time that is impractical in experiments. In this work, we focus on non-adiabatic GQGs. In addition, GQG has also been investigated in the context of semiconductor quantum dots using spin states \cite{Solinas.03a,Mousolou.14,Mousolou.17b,Mousolou.18a, Mousolou.18b, Kang.20,Zhang.20} and charge states \cite{Wang.16,Mousolou.17,Mousolou.18a,Wang.18}, based on non-adiabatic evolutions. Nevertheless, detailed implementation of GQGs in charge qubit systems, especially in those driven by microwave at sweet spots, is lacking in the literature. When operating near the detuning sweet spot with a microwave field, the detuning noise can be suppressed effectively. However, the leading order effect of the tunneling noise is, at the same time, amplified. Moreover, comparing to other qubits mentioned above, the relaxation time of a charge qubit in DQD is very short, only about 10 ns \cite{Kim.15}. Therefore, whether GQGs can offer fidelity improvement in a realistic noise environment remains unsettled.

Here, we propose a theoretical framework to implement universal GQGs for charge qubits. 
Arbitrary single-qubit GQGs can be achieved by introducing strongly microwave-driven oscillating gates near the detuning sweet spot. The evolution path in the parameter space is divided into three distinct parts, such that in each part of the evolution the state evolves along the longitude of the Bloch sphere, and the dynamical phase is canceled out \cite{Zhao.17,Tao.18,Xu.19,Zhang.20}. As a result, a pure geometric gate is obtained. Surprisingly, we find that, even though the amplitude of the tunneling noise is normally smaller than the detuning noise \cite{huang.18}, it can still play a critical role for GQGs. While the detuning noise is suppressed by microwave gating near the detuning sweet spot, the tunneling noise, on the other hand, is amplified.
Thus, GQGs have weak dependence on the detuning noise, but is more sensitive to the tunneling noise. By comparing the performance between GQGs and dynamical gates in the realistic noise environment, we show that GQGs can outperform dynamical gates for a wide range of tunneling noise levels. The fidelity improvement afforded by GQGs can reach as high as 30\% if the strength of the detuning noise and the tunneling noise are comparable.

On the other hand, scalable quantum information processing requires coupling between adjacent qubits. Conventionally, the two-qubit gate for the charge qubits are performed using direct capacitive coupling between quantum dots, the fidelity of which is rather low (less than 70\%) due to the strong charge noise \cite{Li.15,macquarrie.20}. Recently, experiments have shown great progresses of imposing strong and long-range coupling between charge qubits via virtual microwave photons of high-impedance SQUID array resonators \cite{Van.18}. It is shown that by introducing the dipole coupling via the resonator to control the detuning value, the charge noise can be suppressed substantially \cite{stockklauser.17,Van.18,abadillo.19,Wang.20,burkard.20}. In these works, the qubit-resonator coupling is performed in the dispersive regime, namely, the coupling strength is much smaller than the energy difference between the qubit and the resonator. In this case, the effective coupling strength is relatively weak, leading to longer gate time and thus low gate fidelities. However, we show that the holonomic entangling gate can be alternatively realized in the resonant coupling regime, where the effective coupling strength will be much larger and the gate can also be protected by the geometric manipulation. In this work, the total Hamiltonian of the hybrid system consisting of the charge qubits and the superconducting resonator can be described by the well-known Tavis-Cummings model \cite{Fink.09}. When both charge qubits and the resonator are in resonance, the Hamiltonian can form an effective three-level $\Lambda$ structure, which can be used to construct a non-Abelian (holonomic) entangling two-qubit gate \cite{hong.18, Egger.19,li.20}. We numerically find that, the gate fidelity depends more sensitively on the relaxation of the qubit than the decay of the resonator. We also compare this geometric entangling gate to its dynamical counterpart. The geometric entangling gate has a fidelity exceeding 95\% according to the noise level in a recent experiment \cite{Van.18}. However, for comparison, the fidelity for the corresponding dynamical gate is only about 80\%. Overall, our results suggest that the charge qubit may greatly benefit from geometric gate operations.

\section{Model}\label{sec:model}

\begin{figure}
	\includegraphics[width=1.0\columnwidth]{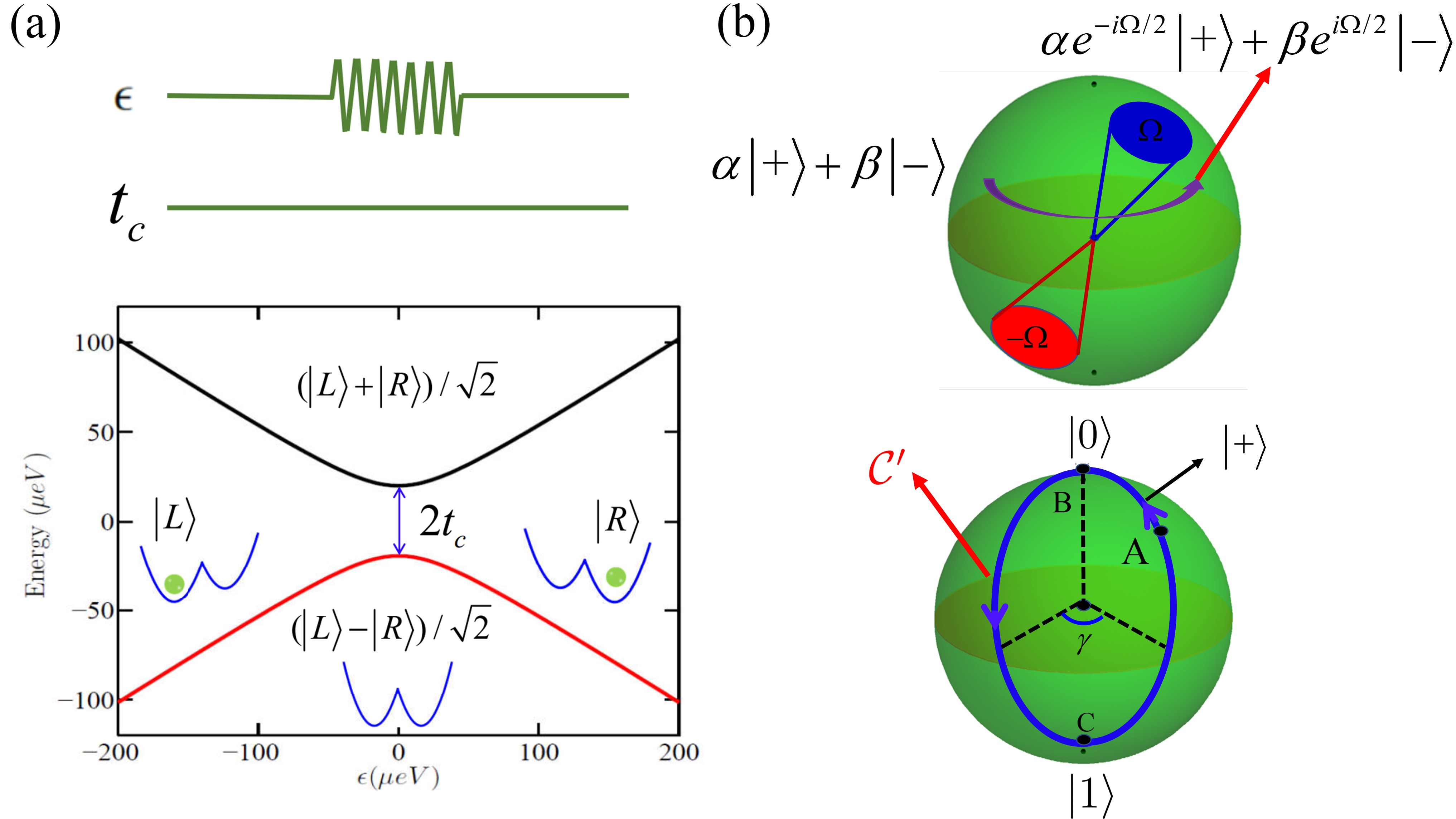}
	\caption{(a) Illustration of the GQG implementation based on charge qubit. The microwave-driven $\epsilon$ pulse oscillates near the sweet spot ($\bar{\epsilon}=0$) while the tunneling is fixed. In the region $\epsilon\gg t_{c}$, the eigenstates are the position states $\left|L\rangle\right.$ and $\left|R\rangle\right.$. Near the sweet spot (namely, $\epsilon \ll t_{c}$), the eigenstates are the asymmetric and symmetric states $\left\{\left|0\rangle\right.= (\left|L\rangle\right. - \left|R\rangle\right.) / \sqrt{2}, \left|1\rangle\right.= (\left|L\rangle\right. + \left|R\rangle\right.) / \sqrt{2} \right\}$ . (b) The schematic of the Abelian GQG. The acquired geometric phase shift of the dressed states is determined by the solid angle enclosed by the closed path. The geometric phase of state $\left|+\right\rangle$ can be achieved through the cyclic evolution along the path A-B-C-A.}
	\label{fig1}
\end{figure}

The charge qubit we consider is shown in Fig.~\ref{fig1}(a). A single electron confined in double quantum dots (DQDs) can occupy either the left (L) or right (R) dot, corresponding to position states labeled by $\left|L\right\rangle$ and $\left|R\right\rangle$ respectively. In absence of noises, the Hamiltonian of this two-level system reads \cite{Van.18}
\begin{equation}
\begin{aligned}
H_{\rm{0}}= t_{c}\tau_{x}-\frac{\epsilon}{2}\tau_{z},
\end{aligned}
\label{eq:Hcq}
\end{equation}
where $\epsilon$ is the detuning value denoting the energy difference between two dots, while $t_{c}$ is the tunnel coupling between the adjacent dots. We set $\hbar=1$ throughout this paper.  Here, $\tau_{x}$ and $\tau_{z}$ are Pauli-X and Z matrices in the position bases $\left\{\left|L\right\rangle, \left|R\right\rangle \right \}$. Instead of using the position bases directly, we define the computational bases as
$\left\{\left|0\right\rangle= (\left|L\right\rangle - \left|R\right\rangle) / \sqrt{2},
\left|1\right\rangle= (\left|L\right\rangle + \left|R\right\rangle) / \sqrt{2} \right\}$ (which are also the eigenstates of $H_{0}$ at the detuning sweet spot $\bar{\epsilon}=0$)
in order to obtain the maximized qubit-resonator coupling strength to facilitate two-qubit operations which will be explained in Sec.~\ref{subsec:couple}.

To take advantage of the detuning sweet spot and realize arbitrary single-qubit gates, qubit manipulation can be performed with oscillating microwave gating of the detuning, i.e., $\epsilon(t) =2A_{\epsilon} \cos(\omega t +\chi)$, where $A_{\epsilon}$ is the amplitude related to the Rabi frequency, $f_{\rm{Rabi}}=A_{\epsilon}/2\pi$. In the computational bases, the strong tunnel coupling $t_{c}$ is analogous to the magnetic field for a single spin-1/2, while the time-dependent $\epsilon(t)$ is similar to a perpendicular oscillating magnetic field, driving state transitions (derivations can be found in Appendix~\ref{appx:Hi}). When $\omega \gg A_{\epsilon}$ and $\omega=2t_c$, the microwave field is in resonance with the energy level splitting at the detuning sweet spot. Neglecting the counter-rotating terms, the effective Hamiltonian in absence of noises and in the frame rotating around $\hat{z}$ with angular frequency $\omega$ can be written as
\begin{equation}
\begin{aligned}
H_{\rm{rot}}(t)=&\frac{A_{\epsilon}}{2}(\cos\chi\sigma_{x}+\sin\chi \sigma_{y}),
\end{aligned}
\label{Hs}
\end{equation}
where $\sigma_{x}$ and $\sigma_{y}$ are Pauli matrices in computational bases. Arbitrary single-qubit control can then be conducted using axes in the $xy$ plane determined by the phase of the microwave field, while the rotating speed is controlled by $A_{\epsilon}$. An entangling two-qubit quantum gate can be achieved by using a charge-qubit-and-resonator hybrid system, as will be explained in Sec.~\ref{subsec:couple}.

\section{Results}
\subsection{Single-qubit GQGs}\label{sec:single}

We first show how the geometric phase is used to construct Abelian single-qubit GQGs. An Abelian GQG takes the form $|k\rangle \rightarrow e^{i f_{k}}|k\rangle$ \cite{Sjoqvist.15}, where $f_{k}$ is the acquired pure geometric phase shift of state $|k\rangle$ after cyclic evolution. Note that, $|k\rangle$ cannot be chosen as $|0\rangle$ or $|1\rangle$ because the resulting gates would commute and thus cannot achieve universal gates \cite{Sjoqvist.15}. Alternatively, we consider $|k\rangle$ as a pair of dressed states involving the computational bases as
\begin{equation}
\begin{aligned}
\left|+\right\rangle&=\cos \frac{\theta}{2}|0\rangle+\sin \frac{\theta}{2} e^{i \varphi}|1\rangle,\\
\left|-\right\rangle&=\sin \frac{\theta}{2} e^{-i \varphi}|0\rangle-\cos \frac{\theta}{2}|1\rangle.
\end{aligned}
\label{eq:spm}
\end{equation}

As seen in Fig.~\ref{fig1}(b), for an arbitrary initial state $\alpha\left|+\right\rangle+\beta\left|-\right\rangle$, the dressed state $\left|+\right\rangle$ ($\left|-\right\rangle$) evolves along a given path on the Bloch sphere. At the final evolution time, it encloses a closed loop (the blue/red circle). After this cyclic evolution, $\left|+\right\rangle$ ($\left|-\right\rangle$) acquires a geometric phase shift: $f_{\pm}=\mp\Omega/2$, where $\Omega$ is the solid angle enclosed by the closed loop. We then have $|\pm\rangle \rightarrow e^{\mp i \Omega/2}|\pm\rangle$, and the final state is thus $\alpha e^{-i \Omega / 2}|+\rangle+\beta e^{i \Omega / 2}|-\rangle$. In the computational bases, this actually defines a geometric transformation \cite{Zhu.02,Zhu.03,Sjoqvist.15}
\begin{equation}
\begin{aligned}
U_{\rm{g}}(\vec{n},\Omega)=e^{-i\Omega/2}\left|+\right\rangle
\left\langle+\right|+e^{i\Omega/2}\left|-\right\rangle\left\langle-\right|=e^{-i\Omega\vec{n}.\vec{\sigma}/2},
\label{eq:Ug}
\end{aligned}
\end{equation}
where $\vec{n}=\left(\sin\theta\cos\varphi, \sin\theta\sin\varphi, \cos\theta\right)$ is a unit vector and $\vec{\sigma}=\left(\sigma_{x},\sigma_{y},\sigma_{z}\right)$. We can see that the dressed states here are exactly the eigenstates of the geometric gate $U_{\rm{g}}(\vec{n},\Omega)$ with eigenvalues $e^{\mp i \Omega/2}$ ($e^{i f_{\pm}}$).
$f_{\pm}$ usually contains two parts: the geometric part and the dynamical part, the latter of which is more sensitive to noises. Therefore, the key to constructing GQGs is to design an appropriate closed path for the dressed states to completely remove dynamical phases, leaving only pure geometric phases.

\begin{figure}
	\includegraphics[width=1.0\columnwidth]{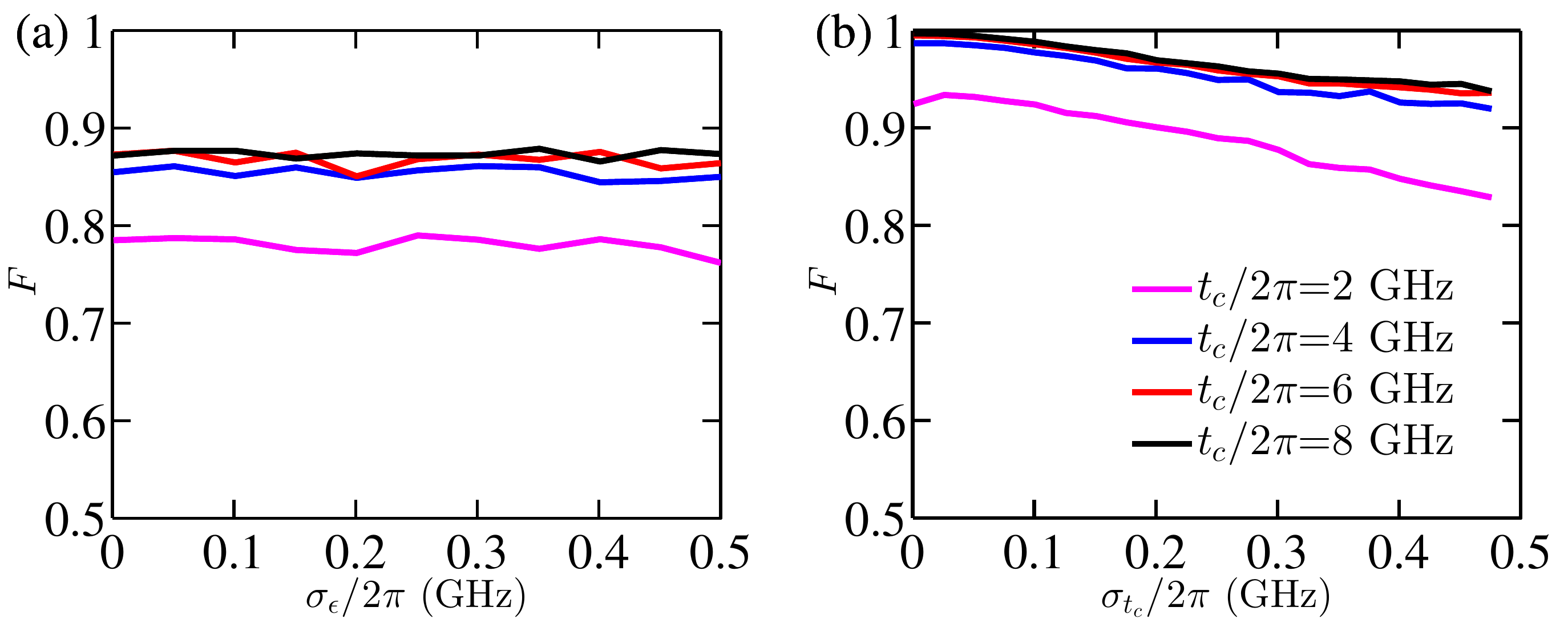}
	\caption{Noise dependence of GQG. (a) The fidelity as a function of the strength of detuning noise $\sigma_{\epsilon}$, where $\sigma_{t_{c}}/2\pi=0.75\ \rm{GHz}$. (b) The fidelity as a function of the strength of the tunneling noise $\sigma_{t_{c}}$, where $\sigma_{\epsilon}/2\pi=0.75\ \rm{GHz}$. For both (a) and (b), the Rabi frequency are fixed as $f_{\mathrm{Rabi}}=A_{\epsilon}/2\pi=\rm{2}\ \rm{GHz}$ , which corresponds to a gate time of $T=0.5$ ns, and the working point is chosen at the detuning sweet spot, i.e., $\bar{\epsilon}=0$.}
	\label{fig2}
\end{figure}

Here, we consider designing the path using the method as introduced in \cite{Zhao.17,Tao.18,Xu.19,Zhang.20}, which is completed in three pieces, forming a closed loop $\mathcal{C'}: t\in [0,T]$ on the Bloch sphere. In Fig.~\ref{fig1}(b), we show the evolution of the dressed state $\left|+\right\rangle$ (the case for $\left|-\right\rangle$ is similar). The evolution of $\left|+\right\rangle$ starts at a given point A, and it travels through the north pole B at time $t=T_1$ via a longitude line and the south pole C at $t=T_2$ via another longitude line, which is $\gamma$ apart from the previous one. Finally, it goes back to the starting point A via the original longitude line. If the evolution from A to B, $U_\mathrm{BA}(T_1,0)=\exp\{-i\int_0^{T_1}H_\mathrm{rot}(t)dt\}$, satisfies
\begin{equation}
\int_{0}^{T_1}A_{\rm{\epsilon}}(t)dt= \theta, \quad \chi_\mathrm{BA}=\varphi-\frac{\pi}{2},
\end{equation}
while the evolution from B to C, $U_\mathrm{CB}(T_2,T_1)$ satisfies
\begin{equation}
\int_{T_1}^{T_2}A_{\rm{\epsilon}}(t)dt= \pi, \quad \chi_\mathrm{CB}=\varphi+\gamma-\frac{\pi}{2},\\
\end{equation}
and similarly the evolution from C to A, $U_\mathrm{AC}(T,T_2)$ satisfies
\begin{equation}
\int_{T_2}^{T}A_{\rm{\epsilon}}(t)dt= \pi-\theta, \quad \chi_\mathrm{AC}=\varphi-\frac{\pi}{2},
\end{equation}
the three connecting pieces together form the desired single-qubit rotation
\begin{equation}
U_{\rm{s}}(\gamma,\theta, \varphi)=U_{\rm{AC}}(T,T_2)U_{\rm{CB}}(T_2,T_1)U_{\rm{BA}}(T_1,0)
=e^{i\gamma\vec{n}.\vec{\sigma}}.
\label{eq:Us}
\end{equation}
Here, parameters $\gamma$, $\theta$ and $\varphi$ are all controlled independently via the microwave field. Therefore, Eq.~\eqref{eq:Us} can constitute an arbitrary rotation on the Bloch sphere. Under the operation of $U_{\rm{s}}(\gamma,\theta, \varphi)$ (at a final time $T$), it is straightforward to find that $U_{\rm{s}}(\gamma,\theta, \varphi)\left|\pm\right\rangle=e^{\pm i\gamma}\left|\pm\right\rangle$, namely, $e^{\pm i \gamma}$ is the eigenvalue of $U_{\rm{s}}(\gamma,\theta, \varphi)$. Thus, $U_{\rm{s}}(\gamma,\theta, \varphi)$ can also be rewritten as
\begin{equation}
\begin{aligned}
U_{\rm{s}}(\gamma,\theta, \varphi)=e^{i\gamma}\left|+\right\rangle\left\langle+\left|+e^{-i \gamma}\right| -\right\rangle\left\langle-\right|,
\label{eq:Uorthor}
\end{aligned}
\end{equation}
using the dressed-state representation. Since the dressed states travel along the longitude all the time, the dynamical phase is cancelled out \cite{Zhao.17,Tao.18,Xu.19,Zhang.20}. Therefore, the global phase factor $\gamma$ is the desired pure geometric phase directly related to the solid angle enclosed by the loop $\mathcal{C'}$, and $U_{\rm{s}}(\gamma,\theta, \varphi)$ represents an arbitrary single-qubit geometric operation. It is worth to note that, the geometric gate time is determined by $\int_{0}^{T}A_{\rm{\epsilon}}(t)dt=2\pi$. Throughout this work, we assume $A_{\rm{\epsilon}}/ 2\pi=2$ GHz according to a recent experiment \cite{Kim.15}, so that the gate time is fixed as $T=0.5$ ns.

Then, we analyze the noise effect on GQGs. Charge noise can cause fluctuations of the electrical potential in the quantum dots. Both the detuning value and the tunneling value are shifted \cite{huang.18} as $\epsilon\rightarrow\bar{\epsilon}+\delta\epsilon$ and $t_{c}\rightarrow \bar{t}_{c}+\delta t_{c}$. Here, $\bar{\epsilon}$ and $\bar{t}_{c}$ indicate the mean values, while $\delta\epsilon$ and $\delta t_{c}$ denote the detuning noise and tunneling noise, respectively. The fluctuation in the energy splitting $E_{01}=\sqrt{{\epsilon^{2}+4 t_{c}^{2}}}$ (which is obtained from Eq.~(\ref{eq:Hcq})) can be analyzed by expanding $E_{01}$ in powers of $\delta\epsilon$ and $\delta t_{c}$:
\begin{eqnarray}\label{eq:expandsion}
E_{01}&=&\sqrt{4 \bar{t}_{c}^{2}+\bar{\epsilon}^{2}}
+\frac{\bar{\epsilon} \delta \epsilon}{\sqrt{4 \bar{t}_{c}^{2} +\bar{\epsilon}^{2}}}\notag\\
&&+\frac{4 \bar{t}_{c} \delta t_{c}}{\sqrt{4 \bar{t}_{c}^{2}+\bar{\epsilon}^{2}}}+O[\delta t_{c}^{2}+\delta\epsilon^{2}].
\end{eqnarray}
The first term in Eq.~(\ref{eq:expandsion}) denotes the energy splitting without noise. The second and the third terms indicate the leading order effects of the detuning noise and tunneling noise, respectively. It is clear that the first order effect of detuning noise can be cancelled out at $\bar{\epsilon}=0$, i.e. the detuning sweet spot ($\partial E_{01} / \partial \epsilon=0$). Furthermore, from the derivation of the effective Hamiltonian in Appendix~\ref{appx:Hi}, we see that the detuning noises in the effective Hamiltonian manifest themselves as the counter-rotating terms ($e^{\pm i \omega t}\delta\epsilon$ or $A_{\epsilon}e^{\pm 2i \omega t}$). Therefore, if we continue to increase the tunneling strength ($\omega=2t_{c}$), its effect can be further suppressed. Similar to the detuning case, there exists a tunneling sweet spot when $t_{c}=0$ ($\partial E_{01} / \partial t_{c}=0$). Normally, detuning noise is assumed to be larger than the tunneling noise in the quantum-dot system \cite{huang.18}. As stated above, to suppress the strong detuning noise and realize universal operation, it is better to apply oscillating microwave gates near the detuning sweet spot with a strong tunnel coupling. On the other hand, this also implies that the tunneling operation is far away from the tunneling sweet spot. From the coefficient of the third term in Eq.~(\ref{eq:expandsion}), one can see that the first order effect of the tunneling noise is maximized when $\bar{\epsilon}=0$. Therefore, when the detuning noise is suppressed by microwave gating near the detuning sweet spot, the tunneling noise becomes dominant. This previously under-appreciated role of the tunneling noise has also been noticed in a recent work on spin decoherence in silicon \cite{huang.18}.

\begin{figure}
	\includegraphics[width=1\columnwidth]{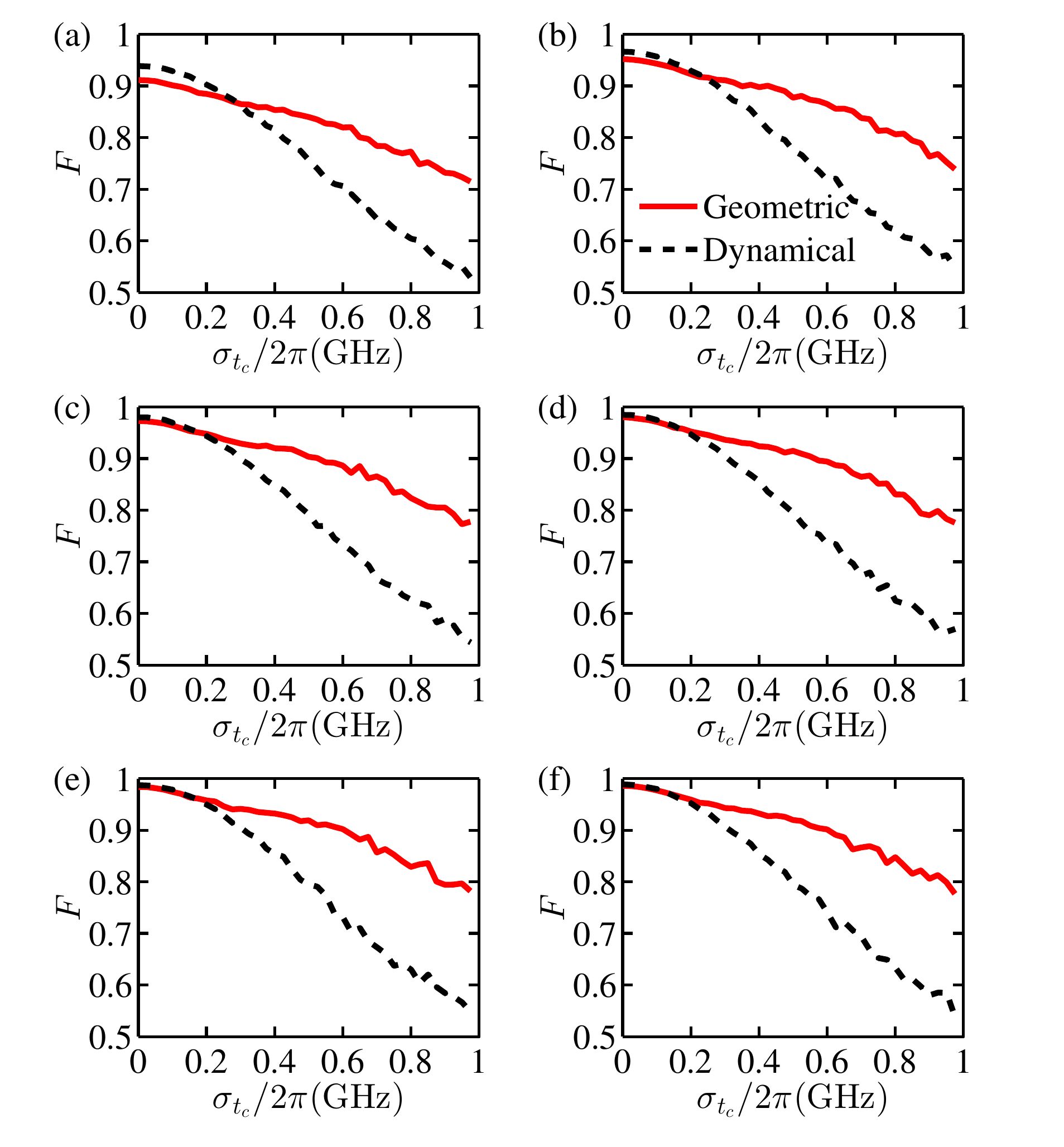}
	\caption{Fidelity of the NOT gate as a function of $\sigma_{t_{c}}$, the strength of the tunneling noise. The dashed black lines denote the results for the dynamical gate and the solid red lines are for the GQGs, respectively. The relaxation times are (a) 10ns, (b) 20 ns, (c) 40 ns, (d) 60 ns, (e) 80 ns, (f) 100 ns. Other related parameters: $f_{\rm{Rabi}}=\rm{2}$ GHz $(T=0.5\ \rm{ns})$, $t_{c}/2\pi=12\ \rm{GHz}$, $\bar{\epsilon}=0$, and $\sigma_{\epsilon}/2\pi=0.75\ \rm{GHz}$.}
	\label{fig:figt}
\end{figure}

To verify this, we numerically plot the fidelity of the NOT gate as an example. The geometric NOT gate is designed using the operator in Eq.~\eqref{eq:Us} as $U_{\rm{s}}(\gamma=-\pi/2,\theta=\pi/2, \varphi=0)$, while the dynamical one is $R(\pi)=\exp\left[-i\frac{\pi}{2}\sigma_{x}\right]$ by setting $\varphi=0$. As shown in Fig.~\ref{fig2}, the fidelity is a function of $\sigma_{\epsilon}$ and $\sigma_{t_{c}}$ for several tunneling values. For simplicity, here we have considered a Gaussian quasistatic noise model \cite{friesen.17}: for each $\sigma_{\epsilon}$ and $\sigma_{t_{c}}$ in our simulation, the detuning noise $\delta\epsilon$ and tunneling noise $\delta t_{c}$ are drawn randomly from normal distributions $\mathcal{N}\left(0, \sigma_{t}^{2}\right)$ and $\mathcal{N}\left(0, \sigma_{\epsilon}^{2}\right)$ respectively, and we average quasistatic noises over 1000 realizations to ensure convergence.
Since the counter-rotating terms in the Hamiltonian are time-dependent, we use standard numerical techniques to solve the master equation. 
Details are described in Appendix~\ref{appx:master}, where the relaxation rate $\gamma_{1}$ is defined. Since here we are focusing on the gate error of the control field, we set the relaxation rate $\gamma_{1}=0$ for simplicity (the relaxation effect will be considered later). In this way, we assume that the infidelity arises from the tunneling and detuning noises, as well as the Rotating Wave Approximation for the counter-rotating terms. We also neglect the control error in $A_{\epsilon}$, which is assumed to be much weaker than the tunneling and detuning noises  \cite{Yang.19a}. In Fig.~\ref{fig2}(a), we show the fidelity results as function of $\sigma_{\epsilon}$ for several tunneling values, where we have assumed $\sigma_{t_{c}}/2\pi=0.75\ \rm{GHz}$. We find that for the smallest tunneling value ($t_{c}/2\pi=2\ \rm{GHz}$), the fidelity is lower than 0.8 for the entire range of $\sigma_{\epsilon}$ considered. Also, we see that results for $t_{c}/2\pi=6\ \rm{GHz}$ (red line) and $t_{c}/2\pi=8\ \rm{GHz}$ (black line) share similar fidelity values of about 85\%. This means the counter-rotating components can be eliminated effectively for large tunneling. In addition, the low fidelity here attributes to the large $\sigma_{t_{c}}$ value. On the other hand, we also find that for all the given tunneling values, the fidelity curves are almost parallel to the $x$ axis. This implies that the geometric gates have weak dependence on the detuning noises. In Fig.~\ref{fig2}(b), we show the fidelity results as functions of $\sigma_{t_{c}}$, assuming $\sigma_{\epsilon}/2\pi=0.75\ \rm{GHz}$. We see that the fidelity shown in Fig.~\ref{fig2}(b) is in general higher than those in Fig.~\ref{fig2}(a). Also, as $\sigma_{t_{c}}$ increases, all the fidelities decrease considerably. This suggests that it is the tunneling noise, rather than the detuning noise, which affects the geometric gates most.

In addition to the detuning and tunneling noises, the fidelity of GQGs may be limited due to the short relaxation time in DQD. To show the superiority of the GQG, we compare the geometric and dynamical NOT gates in the realistic noise environment for six different values of the relaxation time, as shown in Fig.~\ref{fig:figt}. The deviation of the detuning noise $\sigma_{\epsilon}$ at the detuning sweet spot can be calculated via $\sigma_{\epsilon}=c_{\epsilon}\left[2 \ln \left(\sqrt{2 \pi} c_{\epsilon} / \hbar \omega_{l}\right)\right]^{1/2}$ \cite{Yang.19a}, where $\omega_{l}$ is the cutoff frequency and $c_{\epsilon}$ is a parameter controlling the amplitude of the noise. In Ref.~\cite{Yang.19a}, $c_{\epsilon}=0.5\ \mu\mathrm{eV}$ and $\omega_{l} / 2 \pi=1\ \mathrm{Hz}$, such that $\sigma_{\epsilon}\approx 3.12\ \mu\mathrm{eV}$ ($\sigma_{\epsilon}/2\pi\approx0.75\ \rm{GHz}$). In a recent experiment \cite{Kim.15}, the Rabi frequency can be as high as $f_{\rm{Rabi}}=2\ \rm{GHz}$ and the coherence time is about $T_{1}\approx10\ \rm{ns}$ near the detuning sweet spot, which corresponds to a relaxation rate $\gamma_{1}=1/T_{1}\approx0.1\ \rm{GHz}$. Results shown in Fig.~\ref{fig:figt}(a) has a relaxation time consistent to experiments, but we have also performed calculations for other values of relaxation time for completeness, shown in other panels of Fig.~\ref{fig:figt}.
We can see that, for all cases, the dynamical fidelity curves drop faster than those for GQGs. When the tunneling noises are weak ($\sigma_{t_c}/2\pi\lesssim0.3\ \rm{GHz}$), the fidelities of the geometric gates are either lower than (relaxation time $\lesssim$ 20 ns) or comparable (relaxation time longer than 20 ns) to the results from dynamical gates.
We attribute this to the longer evolution time of the GQG, which is $2\pi/A_{\epsilon}$, about twice as long as the dynamical gate. When $\sigma_{t_{c}}$ is large ($\sigma_{t_{c}}/2\pi>0.3\ \rm{GHz}$), the GQG quickly outperforms the dynamical gate and the improvement becomes more pronounced as $\sigma_{t_{c}}$ further increases. This implies that the dynamical gate is more sensitive to the tunneling noise. When $\sigma_{t_{c}}/2\pi=\sigma_{\epsilon}/2\ \pi=0.75$ GHz, the fidelity for the dynamical gate drops to a low value of about 0.6, while for the geometric gate the fidelity is about 0.8, roughly 30\% higher.
Our results indicate that the GQGs can offer substantial improvement in fidelities in a relatively wide range of tunneling noise levels.

\subsection{Two-qubit GQG}\label{subsec:couple}

We consider two DQDs strongly coupled via one of their plunger gates to a high-impedance SQUID array superconducting resonator \cite{Van.18}. For this coupled system, the total Hamiltonian reads
\begin{equation}
H_{\rm{tot}} = H_{\rm{res}}+ \sum_{k=1}^2 H_0^{(k)}+ \sum_{k=1}^2 H_{\mathrm{int}}^{(k)}.
\label{eq:Htot1}
\end{equation}
Here, the resonator term is
$H_{\rm{res}}= \omega_{r}a^{\dagger}a$,
where $\omega_{r}$ denotes the resonant angular frequency of the superconducting resonator, $H_0^{(k)}$ is the $k$th charge qubit (DQD) described by the Hamiltonian in Eq.~(\ref{eq:Hcq}).
The coupling between the resonator and DQD is
$H_\mathrm{int}^{(k)}= g^{(k)}\tau_z^{(k)}(a^{\dagger}+a)$, where $a$ ($a^{\dagger}$) is the bosonic annihilation (creation) operator and $g^{(k)}$ the dipolar coupling strength between the $k$th DQD and the resonator. In order to simulate the Tavis-Cummings Hamiltonian \cite{Fink.09}, we rewrite the Hamiltonian Eq.~\eqref{eq:Htot1} in the eigenbases of the DQD spanned by $\left\{ \left|g\right\rangle,\left|e\right\rangle\right \}$ (where $\left|g\right\rangle$ is the ground state and $\left|e\right\rangle$ is the excited state) as
\begin{equation}
\begin{split}
H_{\rm{tot}}&\simeq
\omega_{r} a^{\dagger}a-\frac{1}{2}\sum_{k=1}^2\omega^{(k)}\tilde{\sigma}_z^{(k)}\\
&+
g^{(k)}\sum_{k=1}^2\sin\eta^{(k)}(a^{\dagger}\tilde{\sigma}_-^{(k)}+\mathrm{H.c.}).
\label{eq:Htot2}
\end{split}\end{equation}
where $\omega^{(k)}=\sqrt{\left(2t_{c}^{(k)}\right)^{2}+\left(\epsilon^{(k)}\right) ^{2}}$, $\tan\eta^{(k)}=2t_{c}^{(k)} / \epsilon^{(k)} $ and $\sin\eta^{(k)}=2 t_{c}^{(k)} / \omega^{(k)}$. Here, we have applied the Rotating Wave Approximation, so the energy-non-conserving terms $a\tilde{\sigma}_-^{(k)}$ and $a^{\dagger}\tilde{\sigma}_+^{(k)}$ are neglected. This is reasonable when $\left|\Delta^{(k)}\right| \ll \omega_{r}+\omega^{(k)}$ \cite{srinivasa.16}, where $\Delta^{(k)}=\omega_{r}-\omega^{(k)}$ is defined as the qubit-resonator detuning. Note that, the Pauli matrices here are $\tilde{\sigma}_z^{(k)}=\left|g\right\rangle \left\langle g\right|-\left|e\right\rangle \left\langle e\right|$, $\tilde{\sigma}_-^{(k)}=\left|g\right\rangle\left\langle e\right|$ and $\tilde{\sigma}_+^{(k)}=\left|e\right\rangle\left\langle g\right|$. We also note that the effective coupling strength is $g^{(k)}\sin\eta^{(k)}$.
Experimentally, $g^{(k)}$ is difficult to change and can be assumed to be a constant.
On the other hand, $\sin\eta^{(k)}$ is determined by the qubit parameters $\epsilon^{(k)} $ and $t_{c}^{(k)}$. In the limit $\epsilon^{(k)}  \gg t_{c}^{(k)} $, the eigenstates of $H_{0}$ are $\left|L\right\rangle$ and $\left|R\right\rangle$, and we have $\sin\eta^{(k)}=2 t_c^{(k)} / \omega^{(k)}\approx2t_c^{(k)}/\epsilon^{(k)} $. In this case, the effective coupling can be very small. Conversely, in the limit $\epsilon^{(k)} \ll t_{c}^{(k)} $, the eigenstates are approximately the computational bases, and we have $\sin\eta^{(k)}\approx1$. As a result, the effective coupling can be maximized. This can be realized by fixing $t_{c}^{(1)}=t_{c}^{(2)}$ and operating $\epsilon^{(k)}$ near the sweet spot, which is similar to the single-qubit case. Below, we set $\sin\eta^{(k)}=1$, and consequently the eigenbases of the DQD ($\left|g\right\rangle$ and $\left|e\right\rangle$) are equivalent to the computational bases ($\left|0\right\rangle$ and $\left|1\right\rangle$), namely, $\tilde{\sigma}=\sigma$.

\begin{figure}
	\includegraphics[width=0.9\columnwidth]{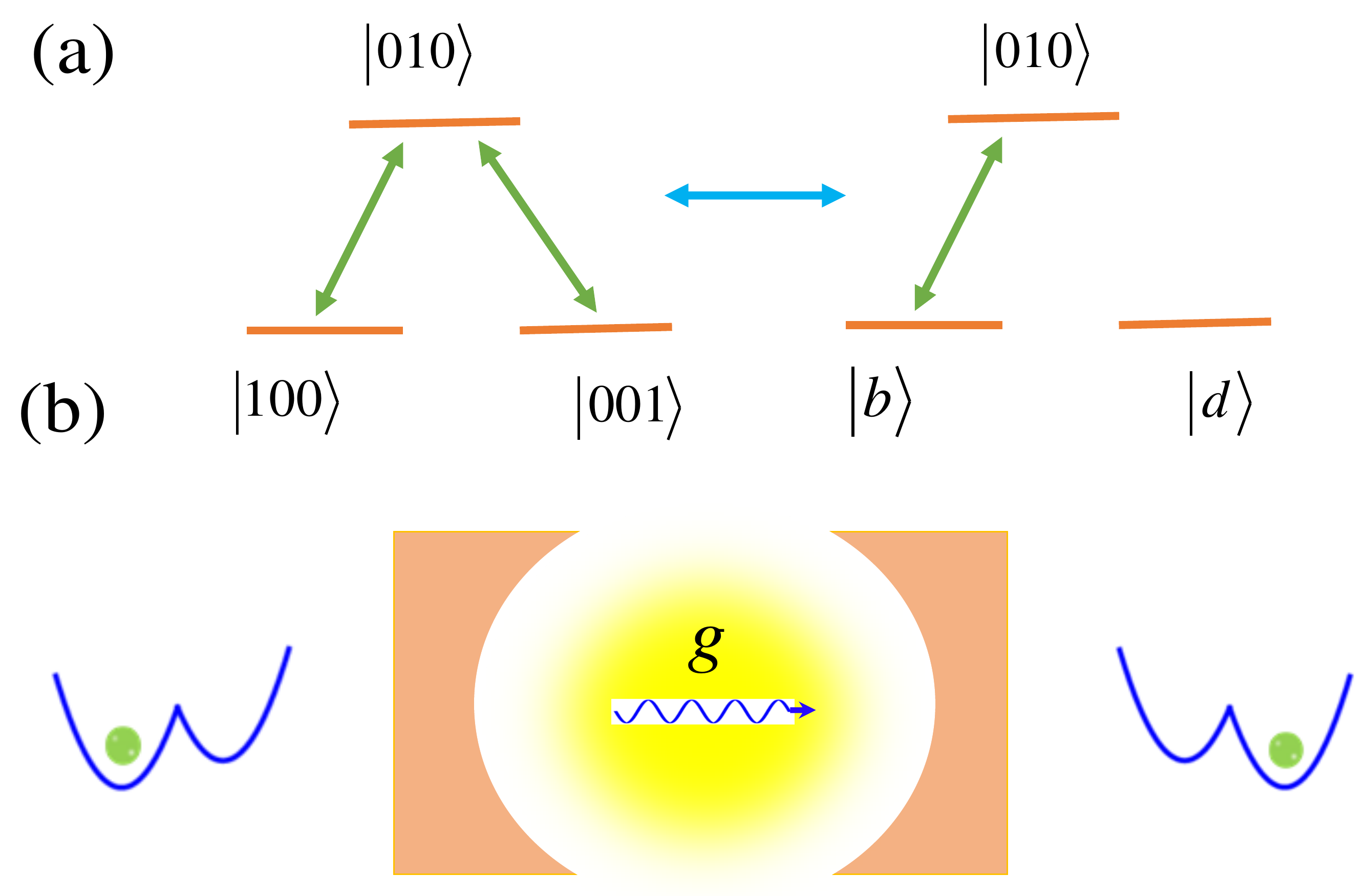}
	\caption{Implementation of the two-qubt GQG. (a) When both qubits and the resonator are in resonance, they form a three-level $\Lambda$ system with the transition between $\left|100\rangle\right.\leftrightarrow\left|010\rangle\right.$ and $\left|001\rangle\right.\leftrightarrow\left|010\rangle\right.$. In the dark and bright representation the dark state decouples the system. (b) Diagram of the charge qubits and resonator coupled system.}
	\label{fig:resonator}
\end{figure}

Next, we transform $H_{\rm{tot}}$ into the rotating frame defined by the resonant frequency $\omega_{r}$. The effective Hamiltonian reads
\begin{equation}
\begin{aligned}
H_{\rm{eff}}=&U^{\dagger}H_{\rm{tot}}U-iU^{\dagger}\frac{\partial U}{\partial t}\\
=&\frac{1}{2}\sum_{k}\Delta\sigma_{z}^{(k)}+\sum_{k}g^{(k)}\left(a^{\dagger}\sigma^{(k)}+\text{h.c}.\right),
\label{Heff}
\end{aligned}
\end{equation}
with
\begin{equation}
\begin{aligned}
U=e^{-i\omega_{r}\left(a^{\dagger}a-\sum_{k}\frac{1}{2}\sigma_{z}^{(k)}\right)t},
\label{eq:Utr}
\end{aligned}
\end{equation}
where we have assumed $\Delta^{(1)}=\Delta^{(2)}=\Delta$. Considering both DQDs and the resonator are resonant, namely, $\Delta=0$, the effective Hamiltonian becomes
\begin{equation}
\begin{aligned}
H_\text{eff}=&\sum_{k}g^{(k)} a^{\dagger}\sigma^{(k)}+\text{h.c}.,
\label{Heff2}
\end{aligned}
\end{equation}
In the single excitation subspace spanned by $\left\{\left|100\rangle\right.,\left|010\rangle\right.,\left|001\rangle\right.\right\}$ (where  $\left|mnq\rangle\right.\equiv\left|m_1\rangle\right.\left|n_r\rangle\right.\left|q_2\rangle\right.$ is denoting the first qubit, the resonator and the second qubit, in an order from left to right), the Hamiltonian in Eq.~(\ref{Heff2}) actually forms a three-level $\Lambda$ system with the transition between $\left|100\rangle\right.\leftrightarrow\left|010\rangle\right.$ and $\left|001\rangle\right.\leftrightarrow\left|010\rangle\right.$, as shown in Fig.~\ref{fig:resonator}(a). This three-level $\Lambda$ system can be alternatively well described by the bright-dark representation \cite{Fleischhauer.96}, with the bright state $|b\rangle$ and the dark state $|d\rangle$ being the superposition states of the lower levels: 
\begin{eqnarray}
\left|b\rangle\right.&=&\sin\frac{\xi}{2}\left|100\rangle\right. -\cos\frac{\xi}{2}\left|001\rangle\right.,\notag \\ \left|d\rangle\right.&=&\cos\frac{\xi}{2}\left|100\rangle\right. +\sin\frac{\xi}{2}\left|001\rangle\right.,
\label{eq:BDs}
\end{eqnarray}
The effective Hamiltonian is therefore
\begin{equation}
\begin{aligned}
H_{\rm{eff}}&=g^{(1)}\left|010\rangle\right.\left\langle 100\right|+g^{(2)}\left|010\rangle\right.\left\langle 001 \right.|+h.c.\\
&=\Omega\left|010\rangle\right.\left\langle b\right.|+h.c.,
\label{eq:Dfs2}
\end{aligned}
\end{equation}
where $\Omega=\sqrt{(g^{(1)})^{2}+(g^{(2)})^{2}}$, $\tan\xi/2=-g^{(1)}/g^{(2)}$. It is clear from Eq.~\eqref{eq:Dfs2} that the dark state $\left|d\rangle\right.$ decouples from the dynamics and the Hamiltonian can be regarded as oscillating between the bright state $\left|b\rangle\right.$ and $\left|010\rangle\right.$.
Thus, $\left|b\rangle\right.$ and $\left|d\rangle\right.$ evolve as
\begin{eqnarray}
\left|\psi_{1}(t)\rangle\right.&=&U_{\rm{eff}}(t)\left|d\rangle\right.=\left|d\rangle\right.\\
\left|\psi_{2}(t)\rangle\right.&=&U_{\rm{eff}}(t)\left|b\rangle\right.=\cos(\Omega t)\left|b\rangle\right.-i\sin(\Omega t)\left|010\rangle\right., \notag
\label{eq:BDs2}
\end{eqnarray}
where $U_{\rm{eff}}=\exp(-i\int_{0}^{T} H_{\rm{eff}}dt)$. When $\Omega T=\pi$ is satisfied, these two states undergo a cyclic evolution with $\left|\psi_{i}(T)\rangle\right.\left\langle \psi_{i}(T)\right|=\left|\psi_{i}(0)\rangle\right.\left\langle \psi_{i}(0)\right|$, $(i=1,2)$. The dark state $\left|d\rangle\right.$ remains the same while the bright state $\left|b\rangle\right.$  acquires a $\pi$ phase factor and thus turns to $-\left|b\rangle\right.$. In the subspace $\left\{\left|d\rangle\right.,\left|b\rangle\right.,\left|010\rangle\right. \right\}$, the evolution operator is
\begin{equation}
U_{\rm{eff}}\left(T\right)=\sum_{i, j=1}^{2}\left[\hat{T} e^{i \int_{0}^{T}\left[A(t)-H_{\rm{eff}}\right] d t}\right]_{i, j}\left|\psi_{i}(0)\right\rangle\left\langle\psi_{j}(0)\right|, 
\end{equation}
with
$A_{ij}(t)=i\left\langle\psi_{i}(t)|\dot{\psi}_{j}(t)\right\rangle$.
Besides, it is easy to find that $\left\langle \psi_i(t)\right|H_{\rm{eff}}\left|\psi_j(t)\rangle\right.=0$ which means there is no transition between $\left|\psi_{1}(t)\rangle\right.$ and $\left|\psi_{2}(t)\rangle\right.$, namely, the parallel-transport condition is satisfied. Thus, $U_{\rm{eff}}(T)$ represents a non-Abelian (holonomic) two-qubit GQG. Furthermore, under the logical basis states $\left\{\left|00\rangle\right., \left|01\rangle\right.,\left|10\rangle\right., \left|11\rangle\right.\right\}$ it takes the form of
\begin{equation}
U_{\rm{eff}}(\xi)=
\begin{pmatrix}
1 &     0    &    0     & 0 \\
0 & \cos\xi  & \sin\xi  & 0\\
0 & \sin\xi  & -\cos\xi & 0 \\
0 &     0    &    0     & -1
\label{eq:Uholo}
\end{pmatrix}
\end{equation}
For the simplest case where $g^{(1)}=g^{(2)}=g_{r}$ such that $\xi=-\pi/2$ we can get a two-qubit entangling gate
\begin{equation}
U_{\rm{ent}}=\begin{pmatrix}
1 &     0    &    0     & 0 \\
0 & 0  & -1  & 0\\
0 & -1  & 0 & 0 \\
0 &     0    &    0     & -1
\end{pmatrix}.
\end{equation}
Note that the negative sign in the elements $\left|11\rangle\right.\left\langle 11\right|$ comes from the evolution of the dual excitation subspace of $\left\{\left|110\rangle\right.,\left|101\rangle\right.,\left|011\rangle\right.\right\}$ \cite{Zhou.18}. That is because in this subspace it can also form another three-level system (see Appendix~\ref{appx:H2}) with the transition between $\left|110\rangle\right.\leftrightarrow\left|101\rangle\right.$ and $\left|011\rangle\right.\leftrightarrow\left|101\rangle\right.$. We can show that $U_{\rm{ent}}$ is actually a perfect entangling gate, namely, it can generate maximally entangled states, see Appendix~\ref{appx:entangle}. Note that the evolution of the Tavis-Cummings Hamiltonian can also be verified by solving the Schrödinger equation by using the probability amplitude method, as seen in Ref.~\cite{Scully.99}.

\begin{figure}
	\includegraphics[width=0.9\columnwidth]{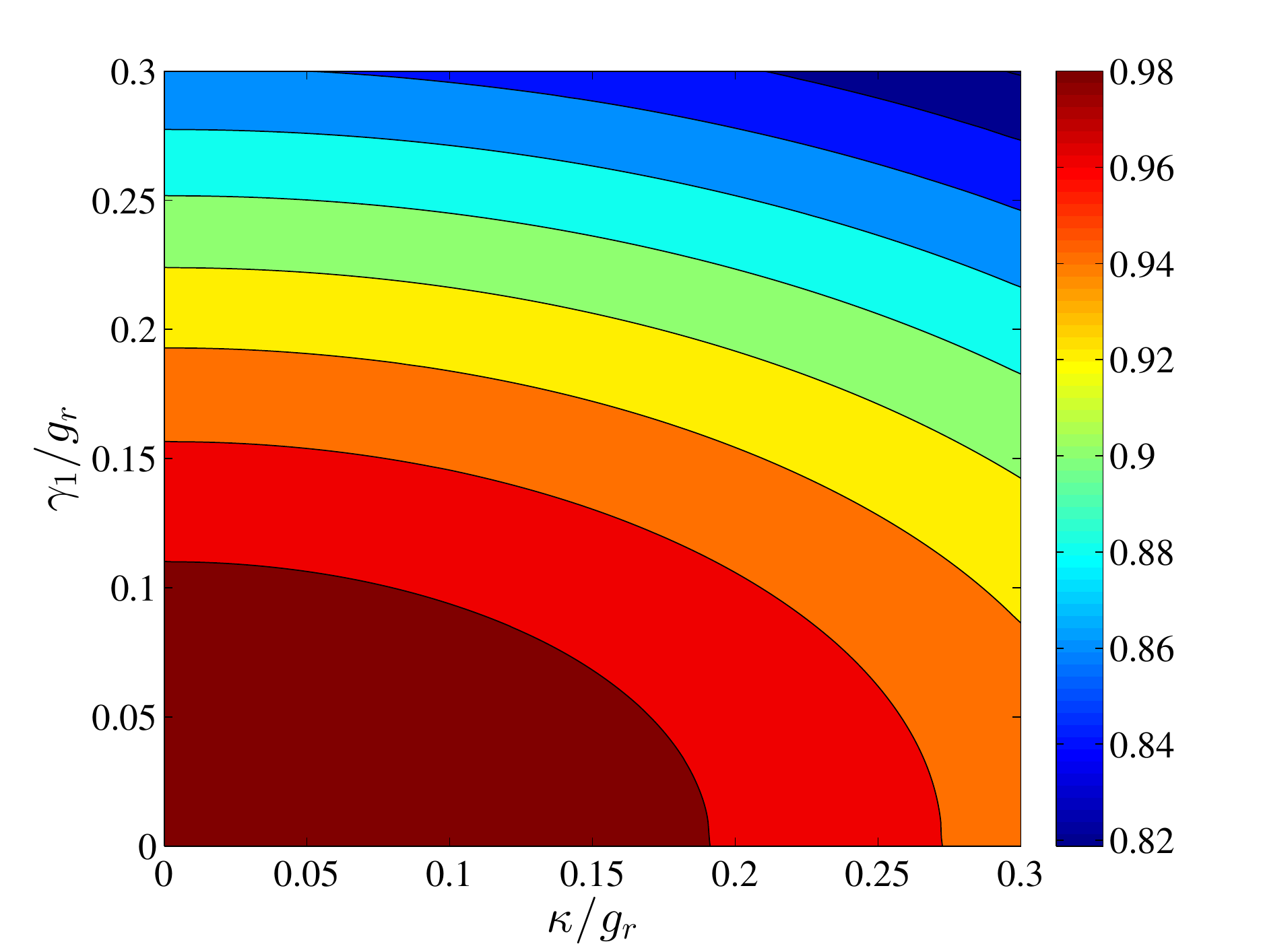}
	\caption{Contour plot of the fidelity of $U_{\rm{ent}}$ as a function of $\kappa/g_{r}$ and $\gamma_{1, k}/g_{r}$. Here, we have assumed $\gamma_{\varphi, k}$=0 and $\gamma_{1, 1}=\gamma_{1, 2}=\gamma_{1}$.}
	\label{fig:contour}
\end{figure}

\begin{figure}
	\includegraphics[width=1.02\columnwidth]{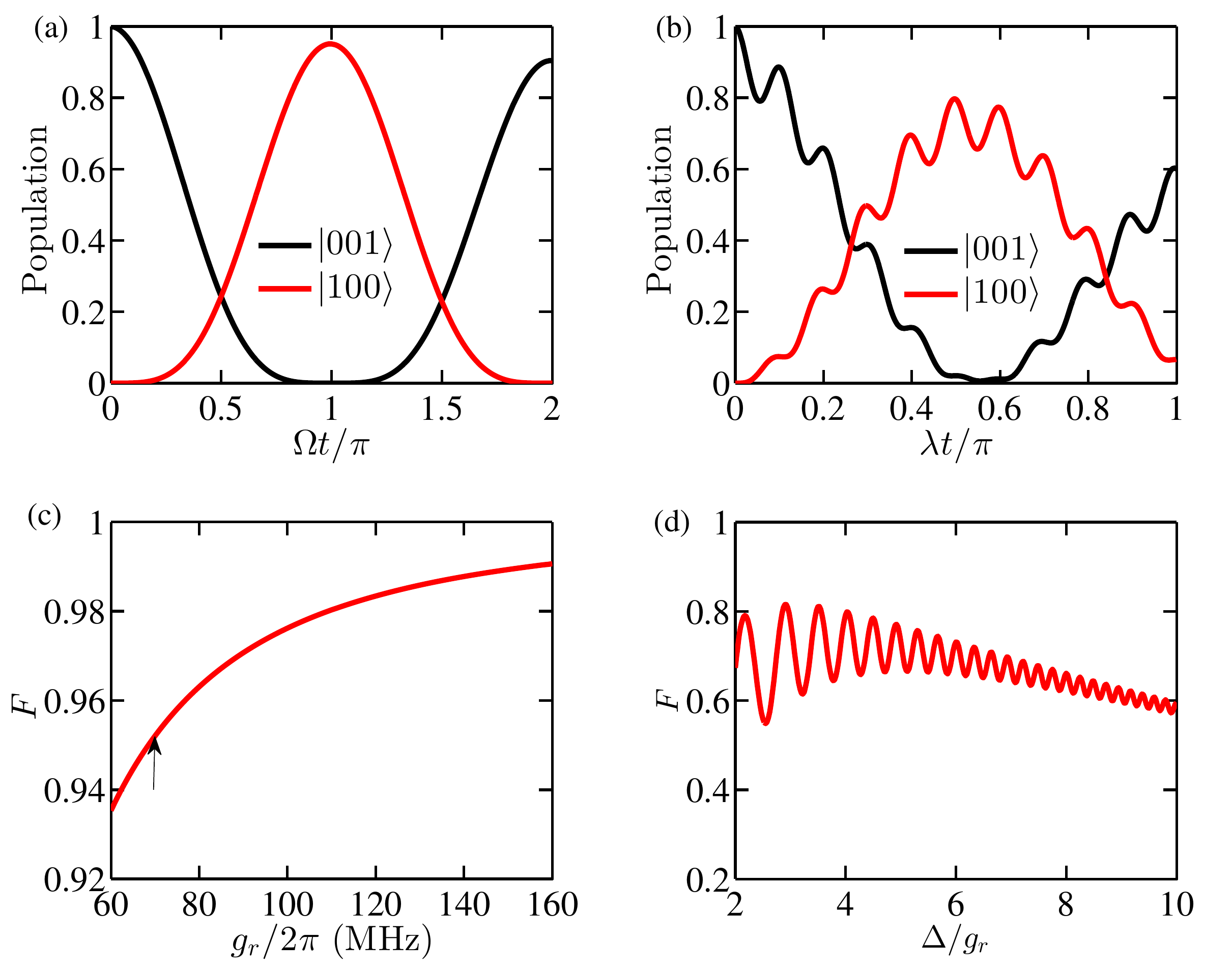}
	\caption{Fidelity and state population of the geometric $U_{\rm{ent}}$ (left column) and its dynamical counterpart $U_{\rm{ent}}^{(2)}$ (right column). (a) State populations of $U_{\rm{ent}}$ versus the evolution time. (b): State populations of $U_{\rm{ent}}^{(2)}$ versus the evolution time. (c) Fidelity of $U_{\rm{ent}}$ as a function of the coupling strength. (d) Fidelity of $U_{\rm{ent}}^{(2)}$ as a function of the ratio $\Delta/g_{r}$.  The simulation parameters is from the recent experiment \cite{Van.18}: for the case $\Delta=0$ [(a) and (c)], $\kappa/2\pi=8.4\ \rm{MHz}$, $\gamma_{1}/2\pi=11\ \rm{MHz}$, while for the case $\Delta\gg g_{r}$ [(b) and (d)], $\kappa/2\pi=12.6\ \rm{MHz}$, $\gamma_{1}/2\pi=8\ \rm{MHz}$. Unless otherwise specified, the coupling strength for all the cases is $g_{r}/2\pi=69\ \rm{MHz}$, the pure dephasing rate $\gamma_{\varphi, k}=0$, and $\Delta/2\pi=280\ \rm{MHz}$.}
	\label{fig:infi}
\end{figure}

We now simulate the performance of this two-qubit entangling gate by numerically solving the master equation as described in Eq.~(\ref{eq:master1}). In Fig.~\ref{fig:contour}, we show contour plot of the gate fidelity as a function of $\gamma_{1}/g_{r}$ and $\kappa/g_{r}$.  $\gamma_{1}$ is the relaxation rate of the qubit while  $\kappa=\omega_{r}/(2Q)$ is the decay rate of the resonator. According to the experiments in Ref.~\cite{Van.18}, distinguishing pure dephasing rate and relaxation rate from the scattering spectrum is difficult. Thus, we absorb the pure dephasing effect into the relaxation rate, and set the pure dephasing rate to zero, i.e. $\gamma_{\varphi, k}=0$. We further assume that the relaxation rate for each qubit is the same, $\gamma_{1, 1}=\gamma_{1, 2}=\gamma_{1}$. We find that, the gate error depends more sensitively on $\gamma_{1}$ than $\kappa$. For example, for the upper bound ($\kappa/g_{r}=0$, $\gamma_{1}/g_{r}=0.3$) the fidelity is about 86\%. However, at the lower bound ($\kappa/g_{r}=0.3$, $\gamma_{1}/g_{r}=0$), the fidelity can reach as high as 94\%. In Fig.~\ref{fig:infi}(a), we show the population results of $U_{\rm{ent}}$. We adopt parameters that are relevant to the recent experiment \cite{Van.18}, where $g_{r}/2\pi=69\ \rm{MHz}$, $\kappa/2\pi=8.4\ \rm{MHz}$ and $\gamma_{1}/2\pi=11\ \rm{MHz}$, which is sufficient to simulate the non-radiative losses and dephasing processes.
We assume the initial state of the coupled system as $\left|001\right\rangle$, i.e. $\left|g,0,e\right\rangle$. Under the operation of $U_{\rm{ent}}$, it is expected to transform to $\left|100\right\rangle$ in the ideal case. We find that the population for $\left|100\right\rangle$ is 95.07\% when $\Omega t=\pi$, which is high relative to the latest experimental results without using GQGs \cite{macquarrie.20}. The corresponding fidelity has been pointed out by the black arrow as shown in Fig.~\ref{fig:infi}(c). To study the dependence on the coupling strength, we further plot the fidelity as a function of the coupling strength $g_{r}$ while keeping other experimental parameters unchanged. As shown in Fig.~\ref{fig:infi}(c), the fidelity increases as $g_{r}$ becomes large. According to the latest experiment \cite{stockklauser.17}, the coupling strength can be as high as $g_{r}/2\pi\simeq160\ \rm{MHz}$, leading to a relatively high fidelity of 99.06\%.

On the other hand, when the qubit and the resonator is operated in the dispersive regime \cite{zheng.00}, i.e. $\Delta\gg g_{r}$, one can also obtain $U_{\rm{ent}}$ in a dynamical way, i.e. $U_{\rm{ent}}^{(2)}$, see Appendix~\ref{appx:iswap}. Since the effective coupling strength is small in the dispersive regime ($\propto1/\Delta$), the corresponding evolution time is  $\sqrt{2}\Delta/g_{r}$ times longer than the resonant case, which consequently exposes the dynamical two-qubit gate $U_{\rm{ent}}^{(2)}$ to more noises. In Fig.~\ref{fig:infi}(b) where $\Delta/2\pi=280\ \rm{MHz}$, we observe that the population result is only 80\% which is much lower than the one of the geometric case. In Fig.~\ref{fig:infi}(d), we further plot the  fidelity of the dynamical gate $U_{\rm{ent}}^{(2)}$ as a function of $\Delta/g_{r}$. We can see that, the fidelity reaches its maximum value about 80\% at  $\Delta/g_{r}\simeq3$ ($\Delta/2\pi\simeq210$ MHz), and decreases as $\Delta/g_{r}$ becomes large due to the overly small effective coupling. When $\Delta/g_{r}=10$  ($\Delta/2\pi\simeq700$ MHz), the fidelity drops to less than 60\%. Comparing Fig.~\ref{fig:infi}(c) and (d),  we conclude that the geometric two-qubit gate is superior to the dynamical gate, offering considerable improvement for two-qubit operation in realistic noise environments. Note that, our holonomic gate is not only suited for the charge qubit but can also be extended to other qubits if the qubits are transversely coupled to the resonator, such as the singlet-triplet qubits \cite{harvey.18}, the resonant exchange qubits \cite{ruskov.19}.

\section{Conclusion}

In conclusion, we have proposed the implementation of universal GQGs for a charge qubit defined in DQDs. We find that, when the charge qubit is operated near the detuning sweet spot, the single-qubit geometric gate is prone to the tunneling noise. However, GQGs can outperform the dynamical gates in a rather wide range of tunneling noise levels, making it particularly suitable to be implemented in this system. In addition, we have designed a non-Abelian (holonomic) entangling two-qubit gate. By solving the master equation we have found that the relevant fidelity is above 95\% for noise relevant to experiments. Therefore, our results offer an alternative yet powerful way to achieve high-fidelity universal geometric quantum computation for the charge qubit.

\section{Acknowledgements}
We thank Bao-Jie Liu for useful discussion. This work was supported by the Key-Area Research and Development Program of GuangDong Province  (No. 2018B030326001), the National Natural Science Foundation of China (Nos. 11905065, 11874156, 11874312), China Postdoctoral Science Foundation (No. 2019M652928),  the Research Grants Council of Hong Kong (No. CityU 11303617),  the Guangdong Innovative and Entrepreneurial Research Team Program (No. 2016ZT06D348), and the Science and Technology Program of Guangzhou (No. 2019050001).

\appendix

\section{Effective Hamiltonian and noise model for the single-qubit gate}\label{appx:Hi}

For a microwave-driven charge qubit, the charge noise couples to the qubit by shifting both the detuning $\epsilon$ and tunneling $t_c$. The Hamiltonian of the coupled system in the basis states spanned by $\left\{\left|L\right\rangle , \left|R\right\rangle \right \}$ can be described by
\begin{equation}
\begin{aligned}
H_{\rm{0}} &= (t_{c}+\delta t_{c})\tau_{x}-\frac{\epsilon+\delta \epsilon}{2}\tau_{z}.
\end{aligned}
\label{eq:H0}
\end{equation}
Under the computational basis states $H_{\rm{0}}$ can be written as
\begin{equation}
\begin{aligned}
H_{\rm{c}} &= \left(\begin{array}{cc}
t_{c}+\delta t_{c} & \frac{\epsilon+\delta \epsilon}{2} \\
\frac{\epsilon+\delta \epsilon}{2} & -t_{c}-\delta t_{c}
\end{array}\right).
\end{aligned}
\label{eq:Hc}
\end{equation}
Here, we consider a microwave-driven operation on the detuning near the sweet spot $\epsilon(t)=2A_{\epsilon}\mathrm{cos}(\omega t+\chi)$. Note that, here we have neglected the error for $A_{\epsilon}$, as it is  much weaker than the tunneling  and the detuning noises considered  \cite{Yang.19a}. Then, we further transform $H_{\rm{c}}$ into the rotating frame at the microwave-field frequency $\omega$:
\begin{widetext}
\begin{equation}
\begin{aligned}
H_{\rm{i}} &= U_{\rm{i}}^{\dagger}H_{\rm{c}}U_{\rm{i}}-iU_{\rm{i}}^{\dagger}\frac{\partial U_{\rm{i}}}{\partial t}
\\
=&\left(\begin{array}{cc}
t_{c}+\delta t_{c}-\frac{\omega}{2} & \frac{A_{\epsilon}}{2}  e^{-i \chi} + \frac{A_{\epsilon}}{2} e^{i(2\omega t+\chi)}+\frac{\delta\epsilon}{2} e^{i\omega t} \\
\frac{A_{\epsilon}}{2}  e^{i \chi} + \frac{A_{\epsilon}}{2} e^{-i(2\omega t+\chi)}+\frac{\delta\epsilon}{2} e^{-i\omega t} & -t_{c}-\delta t_{c}+\frac{\omega}{2}
\end{array}\right),
\end{aligned}
\label{eq:Hi}
\end{equation}
\end{widetext}
where $U_{\rm{i}}=e^{ -i \frac{\omega t}{2} \sigma_{z}}$. For an ideal case, namely, without noise effects and $\omega \gg  \delta\epsilon,A_{\epsilon}$ and $\omega=2t_{c}$ we have
\begin{equation}
\begin{aligned}
H_{\rm{i}}' &= \left(\begin{array}{cc}
0 & \frac{A_{\epsilon}}{2}  e^{-i \chi} \\
\frac{A_{\epsilon}}{2}  e^{i \chi} & 0
\end{array}\right),
\end{aligned}
\label{eq:Hrot}
\end{equation}
where the high-frequency oscillating terms $e^{\pm i\omega t}$ and $e^{\pm 2i\omega t}$ and the detuning noise $\delta\epsilon$ and tunneling noise $\delta t_{c}$ have been neglected.

\section{Master equation}\label{appx:master}

For the single-qubit case, we simulate the geometric quantum gate performance under decoherence by numerically solving the Lindblad master equation \cite{Erik.12}
\begin{eqnarray}
\dot{\rho}=-i\left[H_{\rm{i}}, \rho\right]+\gamma_{1}\mathcal{D}[\hat{L}_{1}] \rho,
\label{eq:mastersingle}
\end{eqnarray}
where
\begin{eqnarray}
\mathcal{D}[\hat{L}_{1}] \rho=\left(2 L_{1} \rho L_{1}^{\dagger}-L_{1}^{\dagger} L_{1} \rho-\rho L_{1}^{\dagger} L\right) / 2
\label{eq:masterL}
\end{eqnarray}
is called the Lindbladian \cite{blais.07}, which describes the effect of the baths on the system, and $\rho$ is the density matrix. Here, we consider $L_{1}=|0\rangle\langle 1|$ with the relaxation rate denoted by $\gamma_{1}$. Note that, here we are using the computational basis to describe the relaxation effect rather than the position basis. The reason is that at the detuning sweet spot ($\bar{\epsilon}=0$), the eigenstates are just the computational basis states, and the relaxation process corresponds to the transition from the higher energy state ($|1\rangle$) to the lower energy state ($|0\rangle$). On the other hand, this relaxation time can be measured by driving the detuning value to the position state (readout point) \cite{Kim.15}. From Eqs. (\ref{eq:Hi}) and (\ref{eq:mastersingle}), we see that the error sources including those arising from the rotating-wave approximation, the charge detuning noise and tunneling noises, as well as the relaxation effect. Then,  the gate fidelity is numerically calculated via \cite{srinivasa.16}
\begin{eqnarray}
F(t)=\mathrm{Tr}\left[\rho_\mathrm{id}(t)\rho(t)\right],
\label{eq:fidelity}
\end{eqnarray}
where $\rho(t)$ is the actual density matrix evolving with time and $\rho_\mathrm{id}(t)$ is the ideal density matrix. Here, we consider the initial state as $|0\rangle$. Then, at the final evolution time $T$, one can obtain $\rho(t)$ by solving the master equation and then calculate the fidelity numerically.

For the two-qubit case, the non-radiative losses and dephasing process of the hybrid quantum-dot-and-resonator system can be described by the master equation \cite{blais.07,wall.07,Van.18}, which is slightly different from the single-qubit case
\begin{eqnarray}
\dot{\rho}=-i\left[H_{\rm{tot}}, \rho\right]+\mathcal{L}_{\mathrm{nr}} \rho,
\label{eq:master1}
\end{eqnarray}
where\begin{equation}
\mathcal{L}_{\mathrm{nr}} \rho=\sum_{k} \left( \gamma_{1, k} \mathcal{D}\left[\sigma_{-}\right] \rho
+  \frac{\gamma_{\varphi, k}}{2}  \mathcal{D}\left[\sigma_{z}\right] \rho \right)
+\kappa \mathcal{D}[a] \rho,
\label{eq:linblad}
\end{equation}
Here, $\gamma_{1, k}$ and $\gamma_{\varphi, k}$ are the relaxation rate and pure dephasing rate for the $k$th DQD, and $\kappa=\omega_{r}/(2Q)$ is the decay rate of the resonator. The resonator has two decay channels, namely the internal channel and the external channel \cite{Van.18,scarlino.21}, where the internal loss of the resonator dominates the external loss. According to the state-of-the-art device the decay rate in the experiment can be as small as $\kappa/2\pi=0.028\ \rm{MHz}$ which corresponds to $Q=10^5$ \cite{Samkharadze.16}.

\section{Effective Hamiltonian for the  qubit-resonator coupled system}\label{appx:H2}
To clearly verify the evolution of the coupled system we write the effective Hamiltonian in the space
spanned by $\left\{ {\left| {100} \right\rangle ,\left| {010} \right\rangle ,\left| {001} \right\rangle ,\left| {110} \right\rangle ,\left| {101} \right\rangle ,\left| {011} \right\rangle ,\left| {000} \right\rangle ,\left| {111} \right\rangle } \right\}$:
\begin{equation}
H_{\rm{eff}}=\begin{aligned}
\left( {\begin{array}{*{20}{c}}
	0&g^{(1)}&0&0&0&0&0&0\\
	g^{(1)}&{\Delta}&g^{(2)}&0&0&0&0&0\\
	0&g^{(2)}&0&0&0&0&0&0\\
	0&0&0&0&g^{(1)}&0&0&0\\
	0&0&0&g^{(1)}&{ - \Delta }&g^{(2)}&0&0\\
	0&0&0&0&g^{(2)}&0&0&0\\
	0&0&0&0&0&0&{\Delta}&0\\
	0&0&0&0&0&0&0&{ - \Delta}
	\end{array}} \right),
\end{aligned}
\label{eq:Hlist}
\end{equation}
When both the two qubits and the resonator are resonant, namely, $\Delta=0$, the Hamiltonian in the subspaces $\left\{ {\left| {100} \right\rangle ,\left| {010} \right\rangle ,\left| {001} \right\rangle } \right\}$ (single excitation) and $\left\{ {\left| {110} \right\rangle ,\left| {101} \right\rangle ,\left| {011} \right\rangle } \right\}$ (dual excitation) forms two effective three-level $\Lambda$ structures.

\section{Verification of $U_\mathrm{ent}$ as a perfect entangling gate}\label{appx:entangle}

The two-qubit gates in the group $SU(4)$ can be classified into two sets operations: local and nonlocal operations. The nonlocal operation can be further divided into a set of perfect entangling gates which can generate the maximally entangled state and another set of nonlocal operations that are not being the perfect entangler. The representative of the former is the CNOT gate, while the latter is the SWAP gate. The nonlocal property of the two-qubit gate $U$ in $SU(4)$ can be described as 	
\begin{equation}	
G_{i}\left(V_{1} \otimes V_{2}\ U\ V_{3} \otimes V_{4}\right)=G_{i}(U)	
\end{equation}	
where $U \in S U(4)$ and $V_{i} \in S U(2)$, while $G_{i}$ is the local invariant conveying the nonlocal properties of $U$. Here, we can see that the operator $U$ actually represents a class of two-qubit gates, and they are equivalent up to local operations in $SU(2)$. The local invariant can be obtained from
calculating the coefficient of the unitary matrix $m(U)$:
\begin{equation}
\begin{aligned}
&G_1=\operatorname{Re}\left[\frac{\operatorname{tr}^{2}[m(U)]}{16}\right],\\
&G_2=\operatorname{Im}\left[\frac{\operatorname{tr}^{2}[m(U)]}{16}\right],\\
&G_{3}=\frac{\operatorname{tr}^{2}[m(U)]-\operatorname{tr}\left[m^{2}(U)\right]}{4}.
\label{G123}
\end{aligned}
\end{equation}
where $m(U)$ is defined as
\begin{equation}
\begin{aligned}
m(U)=\left(Q^{\dagger} U Q\right)^{T} Q^{\dagger} U Q,
\end{aligned}
\end{equation}
and $Q$ is the unitary transformation from the computational basis to the Bell basis:
\begin{equation}
\begin{aligned}
Q=\frac{1}{\sqrt{2}}\left(\begin{array}{cccc}
{1} & {0} & {0} & {i} \\
{0} & {i} & {1} & {0} \\
{0} & {i} & {-1} & {0} \\
{1} & {0} & {0} & {-i}
\end{array}\right).
\end{aligned}
\end{equation}
Whether an operator in $S U(4)$ is being the perfect entangling gate can be well estimated by the conditions with the local invariants \cite{Makhlin.02,Calderon.15}:
\begin{equation}
\begin{aligned}
\sin ^{2} \mu \leqslant 4|G| \leqslant 1,
\label{eq:G1}
\end{aligned}
\end{equation}
and
\begin{equation}
\begin{aligned}
\cos \mu \left(\cos \mu -G_{3}\right) \geqslant 0,
\label{eq:G2}
\end{aligned}
\end{equation}
Here, $G=G_1+i G_2=|G| e^{i \mu }$. For $U_{\rm{ent}}$, it is straightforward to verify $G_1=G_2=0$ and $G_3=-1$, which is the same as the case for the iSWAP gate. Therefore, the conditions in Eqs.~\eqref{eq:G1} and \eqref{eq:G2} are satisfied, and it is equivalent to the iSWAP gate up to the local operations in $SU(2)$.

\section{Derivation of the dynamical two-qubit gate $U_{\rm{ent}}^{(2)}$}\label{appx:iswap}
We transform $H_{\rm{tot}}$ (Eq.~\eqref{eq:Htot2}) into the interaction picture via  $U_{\rm{int}}=e^{-i\left(\omega_{r} a^{\dagger}a-\frac{1}{2}\sum_{k=1}^2\omega^{(k)}\sigma_{z}^{(k)}\right)t}$,
\begin{equation}
\begin{aligned}
H_{\rm{eff}}^{(2)}&=U_{\rm{int}}^{\dagger}g_{r}\sum_{k=1}^2(a^{\dagger}\sigma_-^{(k)}+\mathrm{H.c.})U_{\rm{int}}\\
&=\sum_{k=1}^2g_{r}\left(a^{\dagger}\sigma_-^{(k)}e^{-i\Delta t}+\rm{H.c.}\right).
\label{eq:H2}
\end{aligned}
\end{equation}
In the case $\Delta\gg g_{r}$, there is no energy exchange between the qubit system and the resonator. According to the effective Hamiltonian theory \cite{zheng.00}, the effective Hamiltonian of the hybrid system is
\begin{equation}
\begin{aligned}
H_{\rm{eff}}^{(2)}&=\lambda\sum_{k=1}^2\left(\left|e^{(k)}\right\rangle\left\langle e^{(k)}\right|a a^\dagger-\left| g^{(k)}\right\rangle\left\langle g^{(k)}\right| a^\dagger a\right)\\&+\lambda\left(\sigma_+^{(1)}\sigma_-^{(2)}+\sigma_-^{(1)}\sigma_+^{(2)}\right),
\end{aligned}
\label{eq:H3}
\end{equation}
where the effective coupling strength is $\lambda=g_{r}^{2}/\Delta$. In Eq.~(\ref{eq:H3}), the first and the second terms denote the photon-number dependent Stark shifts, the last term describes the effective coupling of the two qubits induced by the virtual interaction between the qubit and the resonator \cite{zheng.00}. If we assume the resonator is initialized at the zero-photon state, the resonator effect can be further dropped out, and thus  Eq.~(\ref{eq:H3}) reduces to
\begin{equation}
\begin{aligned}
H_{\rm{eff}}^{(2)}&=\lambda\left(\sum_{k=1}^2\left|e^{(k)}\right\rangle\left\langle e^{(k)}\right|+\sigma_+^{(1)}\sigma_-^{(2)}+\sigma_-^{(1)}\sigma_+^{(2)}\right).
\end{aligned}
\label{eq:H4}
\end{equation}
The evolution operator of this Hamiltonian is thus
\begin{equation}
U_{\rm{ent}}^{(2)}(t)=\begin{pmatrix}
1 & 0  & 0 & 0 \\
0 & \frac{1}{2}\left(1+e^{-2 i \lambda t}\right)  & \frac{1}{2}\left(-1+e^{-2 i \lambda t}\right) & 0\\
0 & \frac{1}{2}\left(-1+e^{-2 i \lambda t}\right)  & \frac{1}{2}\left(1+e^{-2 i \lambda t}\right) & 0 \\
0 & 0  & 0 & e^{-2 i \lambda t}
\end{pmatrix}.
\label{eq:Uent2}
\end{equation}
When $\lambda t=\pi/2$, $U_{\rm{ent}}^{(2)}(\frac{\pi}{2\lambda})$ is equivalent to $U_{\rm{ent}}$.

\end{document}